# Conceptual Designs of Dipole Magnet for Muon Collider Storage Ring

I. Novitski, V.V. Kashikhin, N. Mokhov, A.V. Zlobin

*Abstract*— Conceptual designs of a superconducting dipole magnet for a Storage Ring of a Muon Collider with a 1.5 TeV center of mass (c.o.m.) energy and an average luminosity of $10^{34}$ $cm^{-2}s^{-1}$ are presented. In contrast to proton machines, the dipoles for the Muon Collider should be able to handle ~0.5 kW/m of dynamic heat load from the muon beam decays. The magnets are based on $Nb_3Sn$ superconductor and designed to provide an operating field of 10 T in the 20-mm aperture with the critical current margin required for reliable machine operation. The magnet cross-sections were optimized to achieve the best possible field quality in the aperture occupied by beams. The developed mechanical structures provide adequate coil prestress and support at the maximum level of Lorentz forces in the coil. Magnet parameters are reported and compared with the requirements.

## I. INTRODUCTION

A Muon Collider (MC) proposed in 1969 [1] is seen as a promising energy frontier machine for the future of high energy physics [2], [3]. Particle collisions in the Muon Collider will occur through the intersection of two circulating muon beams inside a compact Storage Ring (SR). Requirements and operating conditions for a MC Storage Ring pose significant challenges to superconducting magnets [2], [4]. The dipole magnets should provide a relatively high magnetic field to reduce the Storage Ring circumference and thus maximize the number of muon collisions during their lifetime. Unlike dipoles in proton machines, they should allow the muon decay products to escape the magnet helium volume in order to reduce the heat load on the MC cryogenic system. This imposes additional challenges for the dipole design.

This paper summarizes the results of conceptual design studies of superconducting magnets for the Storage Ring of a Muon Collider with a 1.5 TeV c.o.m. energy and an average luminosity of $10^{34}$ $cm^{-2}s^{-1}$ [5]. These studies included the choice of superconductor and magnet designs to achieve the required field in MC Storage Ring dipole magnets within the specified apertures with appropriate operating margins and accelerator field quality.

Manuscript received August 3, 2010.
This work was supported by Fermi Research Alliance, LLC, under contract No. DE-AC02-07CH11359 with the U.S. Department of Energy.
Authors are with the Fermi National Accelerator Laboratory, Batavia, IL 60510 USA (corresponding author phone: 630-840-8192; fax: 630-840-3369; e-mail: zlobin@fnal.gov).

## II. MAGNET REQUIREMENTS

Muon Collider target parameters are summarized in Table I. The MC Storage Ring lattice and the Interaction Region layout consistent with these parameters are reported in [6].

TABLE I MC STORAGE RING PARAMETERS

| Parameter | Unit | Value |
|---|---|---|
| Beam energy | TeV | 0.75 |
| Nominal dipole field | T | 10 |
| Circumference | km | 2.5 |
| Momentum acceptance | % | ±1.2 |
| Transverse emittance, $\varepsilon_N$ | π·mm·mrad | 25 |
| Number of interaction points | | 2 |
| $\beta^*$ | cm | 1 |

The MC Storage Ring is based on dipole magnets with a nominal field of 10 T. The small transverse beam size ($\sigma$~0.5 mm) requires a small aperture of only ~10 mm in diameter. However, electrons from the muon decays produce a 0.5 kW/m dynamic heat load localized in the horizontal direction predominantly on the inner side of the storage ring. This heat needs to be intercepted outside of the magnet helium vessel.

## III. MAGNET DESIGNS AND PARAMETERS

### A. Strand and Cable

The level of operating magnetic field in MC Storage Ring magnets excludes using traditional NbTi magnets. Recent progress with a new generation of high-field accelerator magnets suggests using $Nb_3Sn$ superconductor, which has the most appropriate combination of the critical parameters $J_c$, $T_c$, $B_{c2}$ and is commercially produced at the present time in long lengths. $Nb_3Sn$ strand and cable parameters used in this study are reported in Table II.

TABLE II STRAND AND CABLE PARAMETERS

| Parameter | Unit | Keystoned cable | Rectangular cable |
|---|---|---|---|
| Number of strands | | 37 | 37 |
| Strand diameter | mm | 0.80 | 0.80 |
| Cable inner thickness | mm | 1.63 | 1.74 |
| Cable outer thickness | mm | 1.84 | 1.74 |
| Cable width | mm | 16.32 | 16.32 |
| Cu/nonCu ratio | | 1.17 | 1.17 |
| $J_c$(12T, 4.2K) | A/mm$^2$ | 2750 | 2750 |

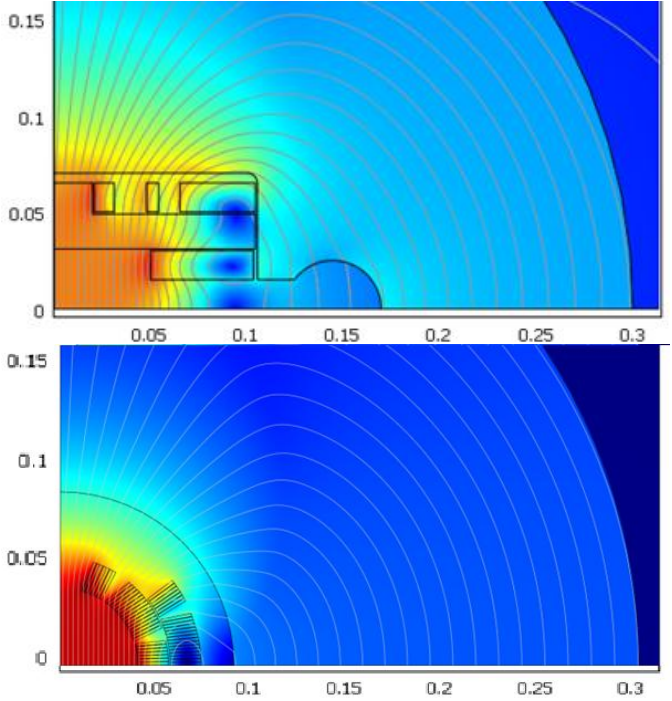

Fig.1. Magnetic field diagrams for MC Storage Ring dipole based on open mid-plane block-type coil (top) or large-aperture shell-type coil (bottom).

*B. Magnet Designs and Parameters*

The dipole requirements and operating conditions call for either an open mid-plane design approach with absorber placed outside the coil or a large-aperture traditional design with absorber inside the magnet aperture [2], [7], [8].

Cross-sections of MC Storage Ring dipoles based on open mid-plane block-type (top) and large-aperture shell-type (bottom) coils are shown in Fig.1. The main magnet parameters are summarized in Table III. Magnetic analysis was performed using ROXIE [9] and OPERA [10] codes.

TABLE III STORAGE RING DIPOLE PARAMETERS

| Parameter | Open mid-plane design | Shell-type design |
|---|---|---|
| $B_{max}$ in coil at 4.5K (T) | 13.5 | 13.7 |
| $B_{max}$ in bore at 4.5 K (T) | 11.2 | 12.5 |
| $B_{op}$ (T) | 10.0 | 10.0 |
| $F_x$ at $B_{op}$ (kN/m) | 3796 | 3033 |
| $F_y$ at $B_{op}$ (kN/m) | -1694 | -1498 |

The open mid-plane block-type dipole uses the rectangular cable described in Table I and a cold yoke with a vertical gap of 170 mm and a horizontal gap of 210 mm for superconducting coils. The yoke also has two large holes in the mid-plane outside the coil. One of these holes contains a ~27 mm thick absorber operating at a temperature of ~70 K. The mid-plane coil-to-coil gap of 30 mm is sufficient to accommodate a beam pipe and provide the required vertical mid-plane open space of at least 10 mm.

The shell-type dipole is based on the keystoned cable described in Table I and a circular cold iron yoke with an inner diameter (ID) of 180 mm. The aperture of the shell-type dipole needs to be large enough to accommodate the ~20 mm diameter beam pipe and ~27 mm thick absorber operating at a temperature of ~70 K, a ~3 mm insulation vacuum space, a ~3 mm magnet cold bore and a ~2 mm inner helium channel. Since the decay particles are localized mainly on one side of the beam pipe, the beam pipe could be shifted in the horizontal direction from the magnet center within the good field region reducing the required coil inner diameter to only 80 mm.

Both magnet designs have nearly the same conductor volume. The number of turns in the open mid-plane magnet is 106, and in the shell-type dipole it is 104. The maximum field in the aperture of the open mid-plane dipole is 11.2 T, which provides a 12% margin with respect to the nominal operating field of 10 T at 4.5 K. The maximum field in the aperture of the shell-type dipole is higher (12.5 T vs. 11.2 T) due to the higher efficiency of this design, and the operating margin of this magnet at the nominal field of 10 T is ~25%.

The coil geometry in both designs was optimized to minimize the geometrical field harmonics in the area with circulating muon beams. In the shell-type design the accelerator field quality (dB/B<$10^{-4}$) is achieved within a 50 mm circle and in the open-midplane design within a 40 mm circle. The good field quality areas and the possible beam pipe positions in both designs are shown in Fig. 2.

Both presented dipole designs have large horizontal and vertical Lorentz force components, which may lead to high stress levels in the coils and large coil deformations. Both components need to be supported by an adequate mechanical support structure to minimize turn motion which may cause magnet quench and field quality degradation.

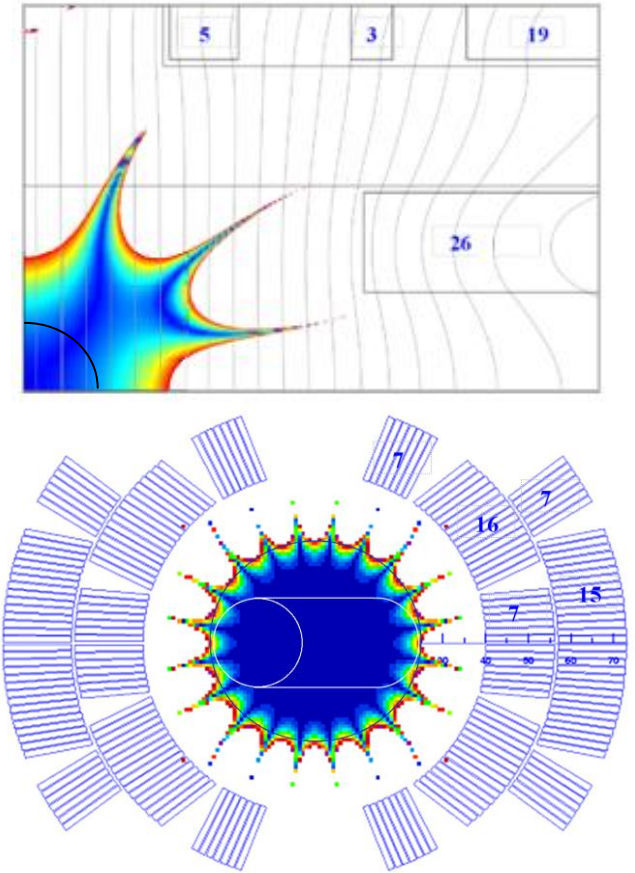

Fig. 2. Field quality in the MC Storage Ring dipole based on open mid-plane (top) or large-aperture (bottom) design. Dark areas correspond to dB/B<$10^{-4}$.

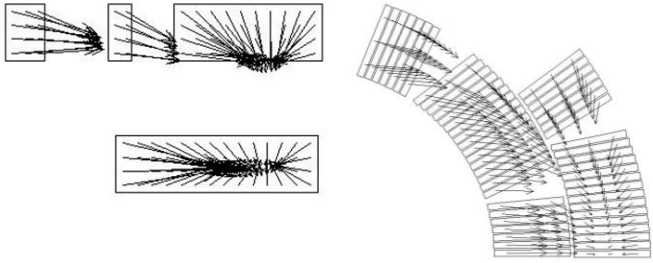

Fig. 3. Lorentz force distributions in the open mid-plane coil (left) and large-aperture shell-type coil (right).

The Lorentz force distributions in the open mid-plane coil and large aperture shell-type coil are shown in Fig. 3. These distributions are quite different, each requiring different approaches to magnet mechanical structures and stress management.

### C. Open Mid-plane Mechanical Design and Analysis

A possible mechanical structure of the open mid-plane dipole described above is shown in Fig. 4. It consists of two double-pancake coils, a vertically split iron yoke, and a thick stainless steel skin with two alignment keys and two control spacers. Two double pancake block-type coils, wound around Ti poles with 22 mm thick interlayer stainless steel plates, are placed inside an Al cage. This cage provides the required vertical coil separation and contains three holes for cooling pipes on one side and a slot for the beam pipe and an escape pass for the decay particles to the absorber placed in one of the two holes in the iron yoke.

Mechanical analysis of this structure demonstrated that some coil support is required in the mid-plane to reduce the coil deformations and stress concentrations. This support element and the beam pipe wall will intercept some decay particles, increasing the heat depositions in the coil and total load on the magnet cold mass. To minimize these effects, the size and material of this support element will be optimized based on mechanical and radiation calculations.

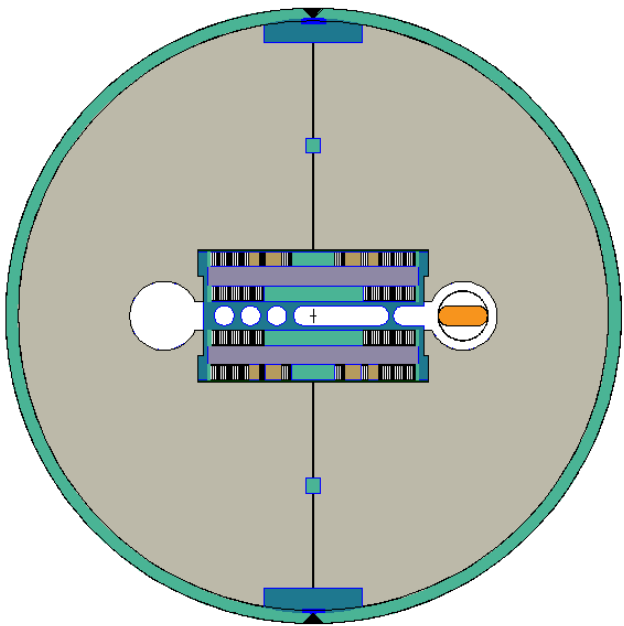

Fig.4. MC Storage Ring open mid-plane dipole based on block-type coil.

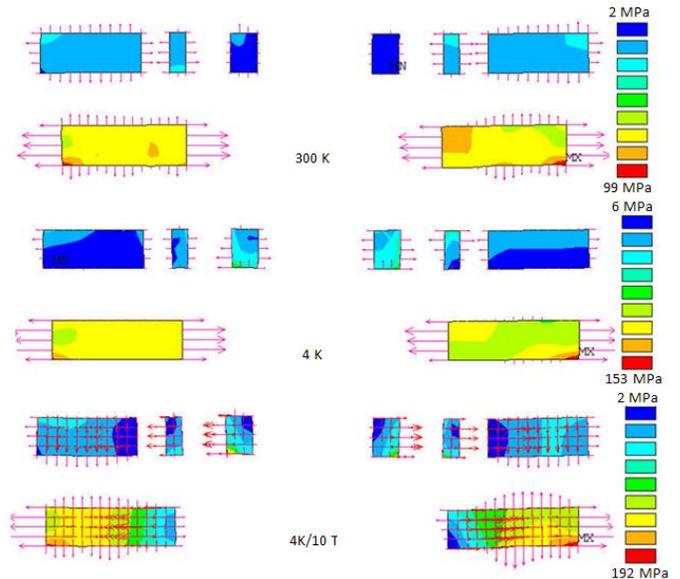

Fig.5. Coil stress distribution in the open mid-plane dipole at room temperature (top), and operating temperature 4.5 K at zero (middle) and at 10 T nominal fields (bottom).

Coil pre-stress is provided by the yoke and skin during welding only in the horizontal direction. The horizontal component of the Lorentz force is supported by the iron yoke and skin. The vertical Lorentz force component is supported by the thick stainless steel plates and Al cage. Both ends of the stainless steel plates are connected to the iron yoke which fixes their vertical position.

Stress distribution diagrams in the open mid-plane dipole coil, calculated using ANSYS at room temperature, after cooling down to 4.5 K and at the magnet nominal operating field of 10 T are shown in Fig. 5. As can be seen, horizontal pre-stress is needed only for the mid-plane coil blocks. The average coil pre-stress at room temperature of ~80 MPa keeps the coil blocks under compression and in contact with the support structure up to the magnet nominal operating field. This is confirmed by the presence of reaction forces on the coil-structure interfaces. Absence of reaction forces on some horizontal interfaces indicates that there is some separation between the coil and support structure. The maximum coil stress is concentrated in the corner of the mid-plane block. It does not exceed 195 MPa, which is acceptable for the $Nb_3Sn$ coils [11] taking also into account that these points are located in low field regions.

Analysis shows that the maximum coil deformations at the nominal operating field do not exceed 60 microns. This seems acceptable with respect to field distortions and coil quenching. Removing the support block in the open mid-plane significantly increases the coil and structure deformations as well as the stress concentrations in the coil. Based on the mechanical analysis, some improvement can be achieved by connecting the stainless steel plates to the iron yoke through the wedges between the coil blocks. However, practical solutions of this problem significantly complicate the support structure, coil fabrication and magnet assembly.

### D. Large-aperture Mechanical Design and Analysis

A mechanical structure of the large-aperture shell-type dipole with shifted beam pipe and internal absorber is shown in Fig. 6. It consists of two double-layer cos-theta coils, vertically split iron yoke, and thick stainless steel skin with two alignment keys and two control spacers. The coils are separated from the iron yoke by thin stainless steel collars. A similar support structure was successfully used in Fermilab's high-field 90-mm $Nb_3Sn$ quadrupole models [12].

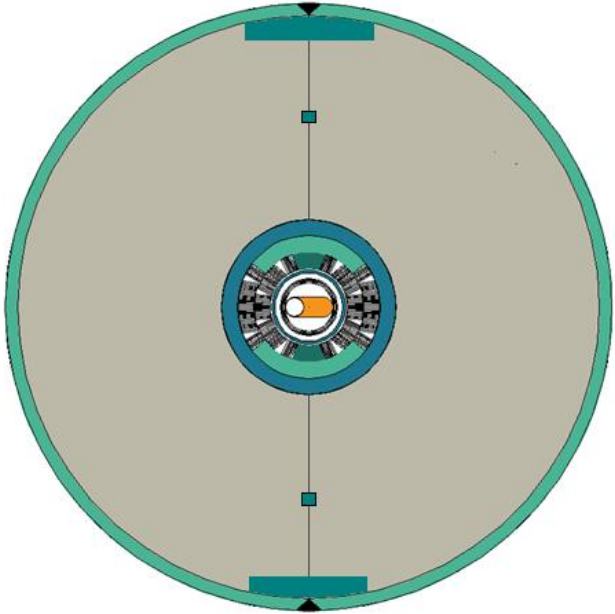

Fig.6. MC Storage Ring large aperture dipole based on shell-type coil.

The initial coil azimuthal pre-stress and geometry control is provided by the collar blocks. The final coil pre-stress and radial support is provided by the yoke and skin.

Stress distribution diagrams for the large-aperture dipole coils calculated using ANSYS at room temperature, after cooling down to 4.5 K, and at the magnet nominal operating field of 10 T are shown in Fig. 7. As can be seen, the average inner-layer coil pre-stress at room temperature of ~120 MPa keeps the coil blocks under compression and in contact with the pole blocks up to the magnet nominal operating field. The maximum coil stress does not exceed 175 MPa. The maximum coil stress could be further reduced by optimizing the radial stress distribution on the pole turns. The maximum coil stress in the shell-type design is lower than in the open mid-plane dipole described above, allowing operation of this magnet type at higher magnetic fields.

## IV. CONCLUSION

Conceptual designs of the $Nb_3Sn$ dipole magnet for the Storage Ring of a Muon Collider with a 1.5 TeV c.o.m. energy and an average luminosity of $10^{34}$ cm$^{-2}$s$^{-1}$ have been developed and analyzed. Two alternative designs, one based on the open mid-plane approach with block type coils and another based on the traditional large aperture cos-theta approach were presented. The magnets are designed to operate at 4.5 K and provide the specified nominal operating field of 10 T with 12-25% margin as well as accelerator field quality in the magnet aperture occupied with muon beams.

Both designs have attractive and difficult features. The open mid-plane dipole concept in principle allows the decay electrons to dissipate their energy in the absorber hidden in the iron yoke and cooled by gas helium or liquid nitrogen. However, the need to support the coils in the gap reduces the efficiency of this approach. Managing the vertical Lorentz force component is the main challenge of the open mid-plane design [13]. Moreover, this magnet design requires an indirect coil cooling scheme which introduces additional complications in magnet design and its operation.

The traditional cos-theta approach requires a large aperture coil which must accommodate both the beam pipe and the absorber. This leads to a large coil volume and large stored energy. However, excellent field quality in the large area inside the coil bore achievable with this design and asymmetric distribution of the heat load from decay electrons in principle allow a significant reduction of the magnet aperture. Meanwhile, the Lorentz force level is noticeably lower and the operating margin is substantially higher in the traditional large-aperture magnet with respect to the open mid-plane magnet. Coil cooling for this design is based on the traditional cooling schemes used in accelerator magnets. Yet, the final coil aperture and the level of heat load in the magnet will depend on the ability to achieve the required field quality in the large volume and the efficiency of the absorber design.

The study and optimization of magnet operating margin, field quality and other key parameters for both designs need to be completed by constructing and testing of a series of magnet models.

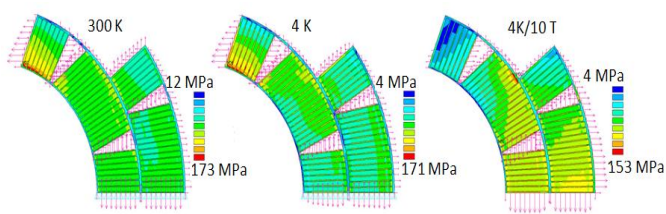

Fig.7. Coil stress distribution in the MC Storage Ring 80-mm dipole at room temperature (left), at 4.5 K zero field (center) and at 10 T field (right).